\documentclass[times, twoside, watermark]{zHenriquesLab-StyleBioRxiv}
\usepackage{bm}
\usepackage{hyperref}
\leadauthor{Slepukhin} 

\begin{document}

\title{Barrier-Free Microhabitats:  Self-Organized Seclusion in Microbial Communities}
\shorttitle{}

\author[a]{Valentin Slepukhin}
\author[b]{V\'ictor Peris Yag\"ue}
\author[a]{Christian Westendorf}
\author[a]{Birgit Koch}
\author[a, c]{Oskar Hallatschek}

\affil[a]{Peter Debye Institute for Soft Matter Physics, Leipzig University, Leipzig 04103, Germany}
\affil[b]{LPENS, Département de Physique, École Normale Supérieure, Université PSL, 75005 Paris, France}
\affil[c]{Departments of Physics and Integrative Biology, University of California, Berkeley, CA 94709}


\maketitle

\begin{abstract}
Bacteria frequently colonize natural microcavities such as gut crypts, plant apoplasts, and soil pores. Recent studies have shown that the physical structure of these spaces plays a crucial role in shaping the stability and resilience of microbial populations~\cite{Karita2022, postek24}. Here, we demonstrate that protected microhabitats can emerge dynamically, even in the absence of physical barriers.  Interactions with surface features  --  such as roughness or friction -- lead microbial populations to self-organize into effectively segregated subpopulations. Our numerical and analytical models reveal that this self-organization persists even when strains have different growth rates, allowing slower-growing strains to avoid competitive exclusion. These findings suggest that emergent spatial structuring can serve as a fundamental mechanism for maintaining microbial diversity, despite selection pressures, competition, and genetic drift.
\end {abstract}

\begin{keywords}
Microbial ecology | Selection | Microfluidics | Self-organization 
\end{keywords}

\begin{corrauthor}
valentin.slepukhin\at uni-leipzig.de
\end{corrauthor}

\section*{Introduction}
Recent microfluidic-based advances in microbial ecology have revealed that bacterial diversity can be surprisingly limited within confined environments. In cavities measuring tens to hundreds of microns, a single microbial strain often dominates, forming a stable population that resists invasion by even faster-growing competitors~\cite{Karita2022}. Once established, this strain persists with minimal diversity, suggesting a form of competitive exclusion, where two species cannot coexist in the same ecological niche~\cite{Gause1934, hardin1960competitive}. These experimental findings align with observations in natural environments such as low internal diversity of microbial community in skin sweat glands ~\cite{conwill2022anatomy} and stability against invasion in the gut environment~\cite{lee2013bacterial, martinez2018experimental, obadia2017probabilistic, lawley2013intestinal}.

However, strain isolation is puzzling when considered in the broader context of the abundant microbial diversity found in natural microbiomes~\cite{donaldson2016gut}. If microbial strains primarily persist through isolation in distinct spatial niches—analogous to how animal species are distributed across different islands~\cite{macarthur1967island, wu1995island}—then direct interactions between strains would be rare. This isolation would preclude essential microbial interactions, such as competition for nutrients, cross-feeding, or mutualistic relationships, which are widely believed to play a crucial role in sustaining diverse microbial communities~\cite{seth2014nutrient, smith2019classification, mataigne2021microbial}. For instance, microbial cooperation within the gut is crucial for various metabolic processes, such as the fermentation of dietary fibers into short-chain fatty acids (SCFAs) like butyrate, which supports colonocyte health \citep{flint2012}, the synthesis of essential vitamins like riboflavin, folate and biotin \citep{leblanc2013, bacher2000}, and cross-feeding interactions where the byproducts of one species, such as acetate, are utilized by others to optimize energy extraction \citep{rakoff2005, louis2014}.

This paradox raises an important question: can microbial strains stably coexist in close proximity without physical barriers, enabling direct chemical interactions? If so, what mechanisms allow such coexistence despite the absence of spatial separation?

In this study, we develop a theoretical framework for self-organized habitat fragmentation in proliferating microbial systems that does not rely on physical barriers. Our model suggests that microbes, through direct mechanical interactions and growth dynamics, can form stable sub-populations even within shared cavities. These sub-populations coexist due to effective isolation driven by two main mechanisms: (i) crowding effects that reduce the effective population to the cells near the closed ends of chambers, and (ii) localized perturbations (defects) and boundary effects that fragment the habitat into distinct niches.

\begin{figure}
\centering
\includegraphics[width=\linewidth]{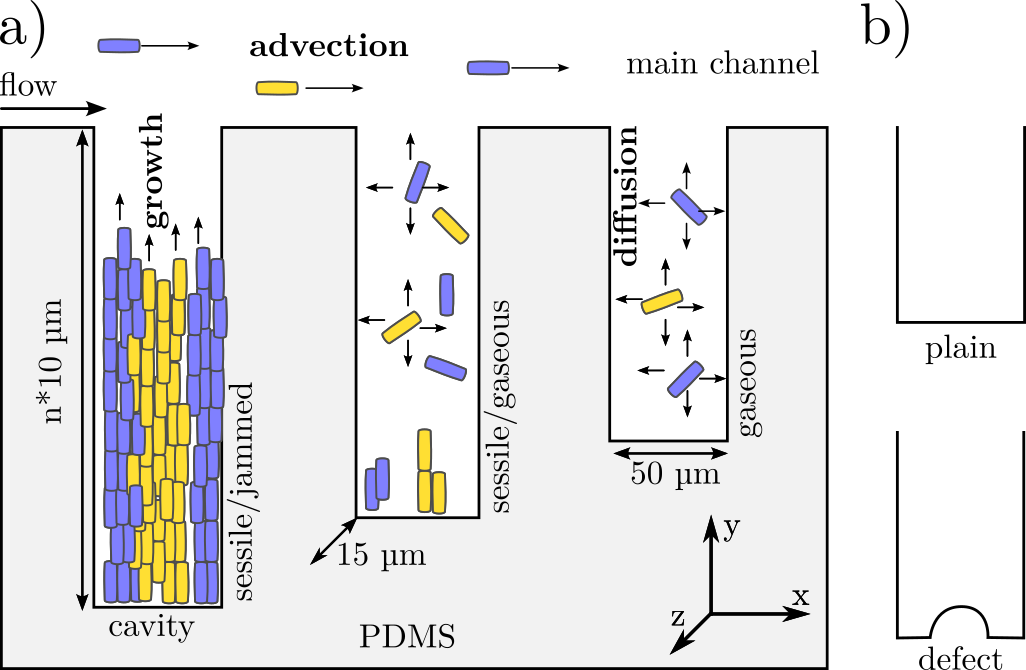}
\caption{Schematic of the microfluidic system under study. a) The device consists of a main channel with an array of varying lengths cavities perpendicular to it, each terminating in a dead-end. Nutrients are supplied continuously through the main channel. Non-motile microbial strains (represented in blue and yellow) are injected into the channel at the beginning of the experiment and subsequently colonize the cavities. As shown previously~\cite{Karita2022} in the longer cavities cells form a jammed population with no extra space available, and in shorter are in a gaseous state, where they can diffuse freely via Brownian motion.   In this study, we examine population dynamics in cavities with smooth floors and cavities featuring a small defect at the floor (panel b).}
\label{fig:intro-fig}
\end{figure}

We demonstrate that this theoretical framework not only explains previously observed effects, such as the impact of structural defects~\cite{postek24}, but also generalizes to other factors like wall friction. To validate our model, we conduct experiments using ``microfluidic panflutes''~\cite{Karita2022} (see Fig.~\ref{fig:intro-fig}) and employ time-lapse microscopy to quantitatively test our predictions for the velocity field of the growing microbial population in the cavities. Our results provide new insights into how microbial communities can self-organize and stably coexist in shared environments, even in the absence of physical separation.

\section*{Results}

\subsection*{Strain fixation and spontaneous demixing controlled by local geometry}

To investigate the microbial population dynamics within the cavity, we employ an individual-based model. In this framework, cells are treated as point-like particles that move according to Langevin dynamics. Instead of computing cell-to-cell interactions explicitly, we approximate their collective effect as a net drift velocity contribution from neighboring cells. As a result, the position $\bm{ x}_i$ of cell $i$ evolves over time according to the Ito-discretized equation:

\begin{equation}  \label{eq:langevin}
     \bm{\dot x}_i = \sqrt{2 D_{\rm self}(\rho)} \cdot \bm{\zeta}_i(t)  + v_{\rm drift}(\rho)
\end{equation}

where $\bm{\zeta}_i(t)$ is unitary uncorrelated Gaussian white noise. The self-diffusivity $D_{\rm self}(\rho)$ depends on the local cell density $\rho$. At very low densities, it aligns with the standard diffusivity observed in Brownian motion, but it approaches zero after the system undergoes the jamming transition.

To derive the expression for the drift velocity, we assume that the net flux of cells follows Fick’s law, given by  $j~=~D_{\rm col}(\rho) \nabla \rho $, where $D_{\rm col}(\rho) $  is the collective diffusivity, modeled based on a hard-disk diffusion framework~\cite{Batchelor_1976}. This collective diffusivity coincides with $D_{\rm self}$ at low densities but diverges to very high values as the system approaches the jamming density\footnote{The validity of Fick’s law at high densities, particularly near the jamming transition, is justified based on comparing it with the proliferating soft disk model~\cite{Karita2022}. }.  By transitioning from the Fokker-Planck equation for density to the Ito-discretized Langevin equation, we obtain: 

\begin{equation}  
     v_{\rm drift}(\rho) = \left( D_{\rm self}(\rho)-D_{\rm col}(\rho) \right) \bm{\nabla}\ln\rho+\bm{\nabla}D_{\rm self}(\rho)
\end{equation}

Cell birth follows a Poisson process with rate $g$. Due to the significantly smaller thickness of the cavity in the $Z$ direction compared to the $X$ and $Y$ directions, we restrict our model to two dimensions (see Fig.~\ref{fig:intro-fig}). Further details on the individual-based model are provided in the SI Section~\ref{note:Note2}.

In this model, we reproduce the transition from an empty to a gaseous, and, finally, to a jammed state observed earlier in \cite{Karita2022}. Interestingly, the density in the gaseous state within the individual-based model is higher than in the continuous limit described in~\cite{Karita2022}, which we attribute to finite-size effects (see SI Fig.~\ref{fig:copmarison_with_comsol}).

\begin{figure}
\centering
\includegraphics[width=\linewidth]{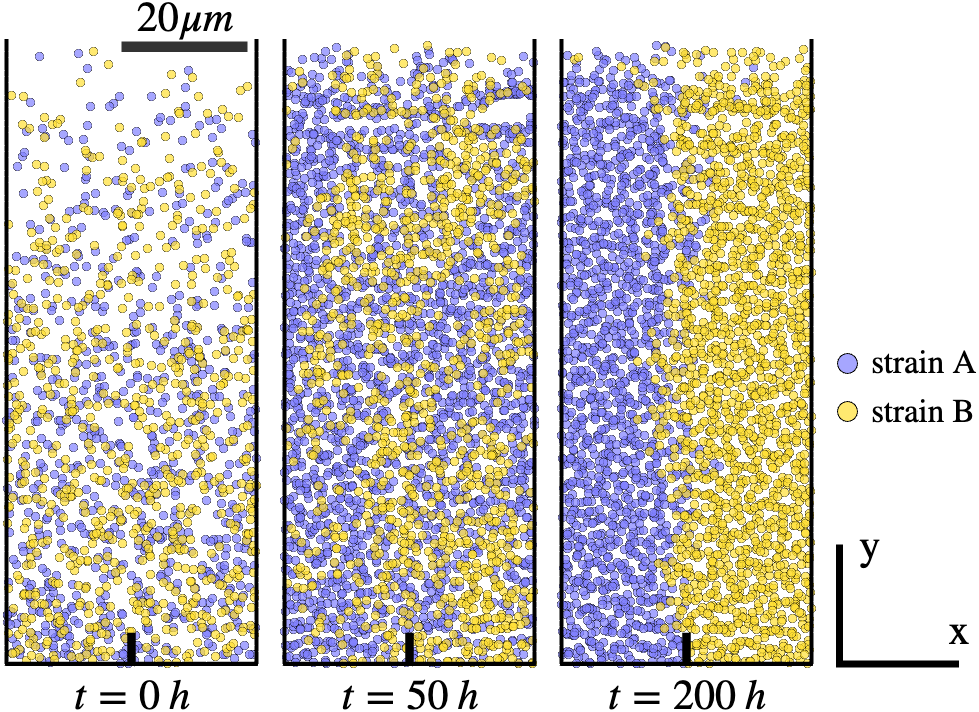}
\caption{Demixing in the jammed population of two neutral strains, individual-based model. The left panel shows the initial state, where both strains are mixed. Over time, random fluctuations cause strain A (blue) to outcompete strain B (yellow) in the left part of the cavity, and strain B to outcompete strain A in the right part of the cavity.  The presence of a small pin-shaped defect (thick black line) at the cavity floor makes these two processes independent from each other, thus allowing a demixed final state in 50\% of all simulations when the left and the right halves of the cavity become dominated by different strains.}
\label{fig:demixing}
\end{figure}

In a system composed of two distinct strains—modeled such that each daughter cell inherits the strain type of its mother — finite-size effects lead to the eventual dominance of one strain over the other. Even if both strains are neutral (i.e., have equal growth rates), one ultimately takes over the entire cavity, thus reaching fixation. This occurs in both the gaseous and jammed states. However, in the jammed state, the exclusion process can be halted by the presence of even a small defect at the cavity floor (see Fig. \ref{fig:demixing}).

In the jammed limit, population dynamics are predominantly governed by cells located at the cavity floor. This situation closely resembles that of expanding populations, where evolutionary dynamics are controlled by cells at the growing frontier~\cite{schreck2023impact, HALLATSCHEK2008158}. In this regime, any cell not positioned at the cavity floor is eventually displaced by the collective flow generated by the population growth. As a result, the bottom layer effectively acts as the active zone driving the dynamics.

A key consequence is that even a small defect along the cavity floor can significantly impact the population structure. Such a defect impedes the lateral movement of cells along the floor, thereby functionally partitioning the cavity into two isolated regions. Despite being negligible in size relative to the total cavity length, the defect effectively creates two independent subcavities in which population dynamics proceed independently. For two neutral strains (i.e. with equal growth rates), different strains will fixate in different subcavities with the probability 50\%. As a result, an initially mixed population can spontaneously demix in the presence of a defect (see Fig.~\ref{fig:demixing} for an exemplary simulation with a pin-shaped defect). 

At the first glance, this outcome may seem counterintuitive, as there is no attractive interaction between identical strains that would drive demixing. Instead, the process arises purely from stochastic fluctuations. The key distinction from non-living matter lies in the proliferation of cells: over time, all cells within a given subcavity will share a common ancestor, i.e. one of the cells originally positioned at the cavity floor. Thus, the separation of strains is not a result of cells actively migrating between subcavities but rather the eventual extinction of all but one lineage within each region.


\subsection*{Defect at the cavity floor balances selective pressure }

If two strains differ in growth rates, the fitter strain generally outcompetes the other finally conquering the entire cavity. To make a quantitative prediction, we derive a coarse-grained, continuous approximation of the individual-based model described in the previous Section.  In this formulation, we treat the density of each strain $\rho_\alpha$  and the corresponding cell flux $j_\alpha$ ($\alpha = A,B $) as continuous variables. By transitioning from the Langevin equation to the Fokker-Planck equation for density, we obtain:
\begin{eqnarray}
    \vec{j}_\alpha = D_{\rm self}(\nabla \rho_{\alpha} - \frac{\rho_\alpha}{\rho} \nabla \rho )+ \frac{\rho_\alpha}{\rho} D_{\rm col} \nabla \rho
    \\
     \frac{\partial \rho_\alpha}{\partial t} + \nabla \cdot \vec{j}_\alpha  = g_\alpha \rho_\alpha
     \label{eq:react-diff}
\end{eqnarray}

\begin{figure}[h!]
\centering
\includegraphics[width=1\columnwidth]{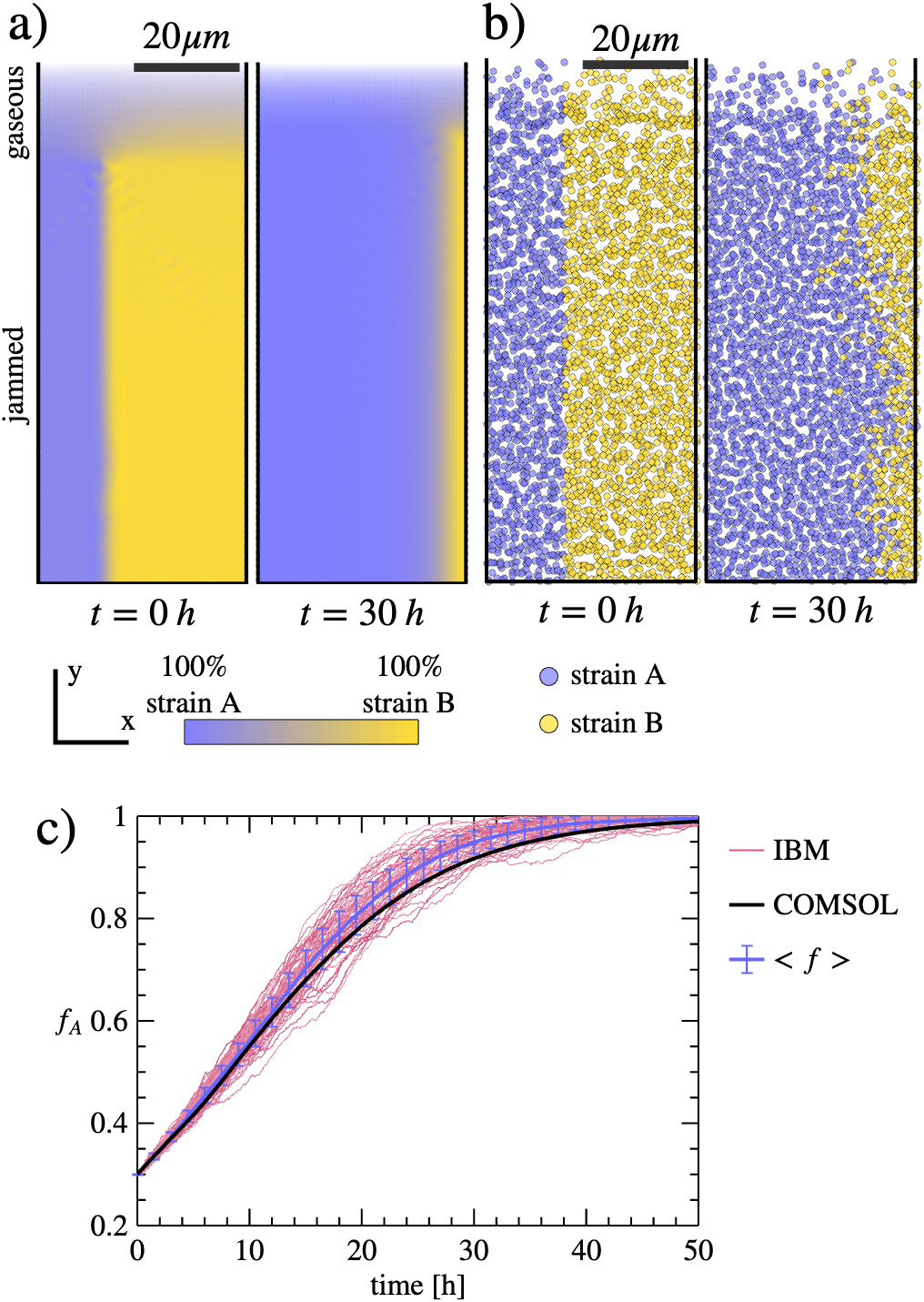}
\caption{ Logistic growth in a cavity with a smooth floor. The two competing strains A and B grow at rates $g_A = 0.33\, h^{-1}$ and $g_B = 0.22\,  h^{-1}$. Diffusion constants are defined in SI Section~\ref{note:Note2}.
(a) Snapshots of the COMSOL simulation defined by Eq.\ref{eq:react-diff}. The color map shows strain frequency: blue for strain A and yellow for strain B. Transparency reflects total cell density, with darker regions corresponding to higher density. White arrows indicate the velocity field.  The population near the cavity opening is in a gaseous state, transitioning to a jammed state toward the dead end of the cavity.
(b) Corresponding snapshots of the individual-based model (IBM) simulation.
(c) Time evolution of strain frequencies described by both the COMSOL and IBM simulations, and further compared to the analytical logistic growth solution (Eq.\ref{eq:logistic}). A single IBM trajectory (red) exhibits stochastic fluctuations, while the average over many runs (light blue) shows a slight deviation from the deterministic logistic profile (black, overlapping with the COMSOL result) due to the earlier extinction of the slower-growing strain.  }
\label{fig:logistic}
\end{figure}

This system of equations is deterministic and, therefore, does not capture the stochastic effects described by the individual-based model. However, if one strain has a significant fitness advantage, stochastic effects primarily act as small corrections to the deterministic process. The continuous model enables us to explore this process analytically and find an explicit functional form for the time dependence of the frequency of the faster-growing strain. Namely, we prove (see SI Sections~\ref{note:Note3},\ref{note:Note4}) that for two strains with growth rates $g_A$ and $g_B$, the frequency  $f_A  $  follows a logistic growth pattern, similar to known results for well-mixed populations~\cite{Moran1962}:

\begin{equation}
    f_A = \frac{1}{1 + e^{-(g_A - g_B) (t - t_{\rm half})}},
    \label{eq:logistic}
\end{equation}

where $t_{\rm half}$ represents the time at which $f_A= f_B = 0.5$. Logistic takeover occurs in gaseous cavities, as well as in jammed cavities in which the initial composition consisted of separated strains arranged in bands along the cavity length. We also show numerically that even if part of the population is in the jammed state, and part is in the gaseous state, the result still holds true (see  Fig.~\ref{fig:logistic}A). In the individual-based model the finite size effects lead to a slight acceleration of the logistic takeover due to accidental extinction of the slow-growing strain (see Fig. \ref{fig:logistic}C).

\begin{figure}
\centering
\includegraphics[width=0.5\linewidth]{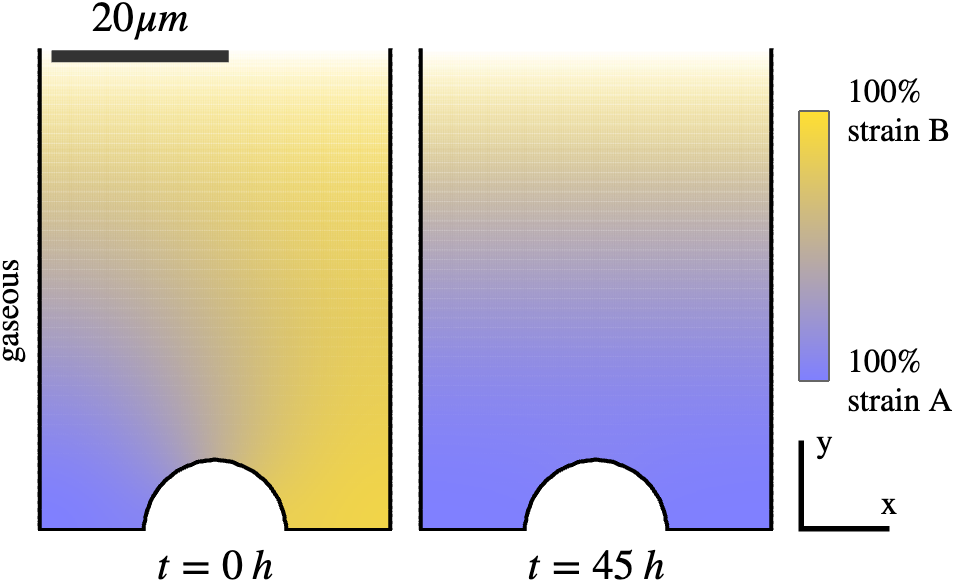}
\caption{COMSOL solution (Eq. \ref{eq:react-diff}) for small cavities only allowing for populations in the   gaseous state. Here the defect on the  floor does not protect a slow growing strain. Left panel: Initial state with strain A (blue), growth rate $g_A = 0.33 h^{-1}$ and strain B (yellow), growth rate $g_B = 0.22 h^{-1} $. Right panel: State at $t = 45 h$ where two strains uniformly mixed. }
\label{fig:gaseous}
\end{figure}

While a simple cavity geometry does not support strain coexistence — leading instead to classical logistic growth or stochastic extinction — modifying the geometry can yield less trivial outcomes. In the gaseous state, geometry plays a minimal role since self-diffusivity remains nonzero, allowing the faster-growing strain to eventually reach any point in the cavity and outcompete the slower-growing strain (see SI Fig.~\ref{fig:gaseous}). 

\begin{figure}
\centering
\includegraphics[width=\linewidth]{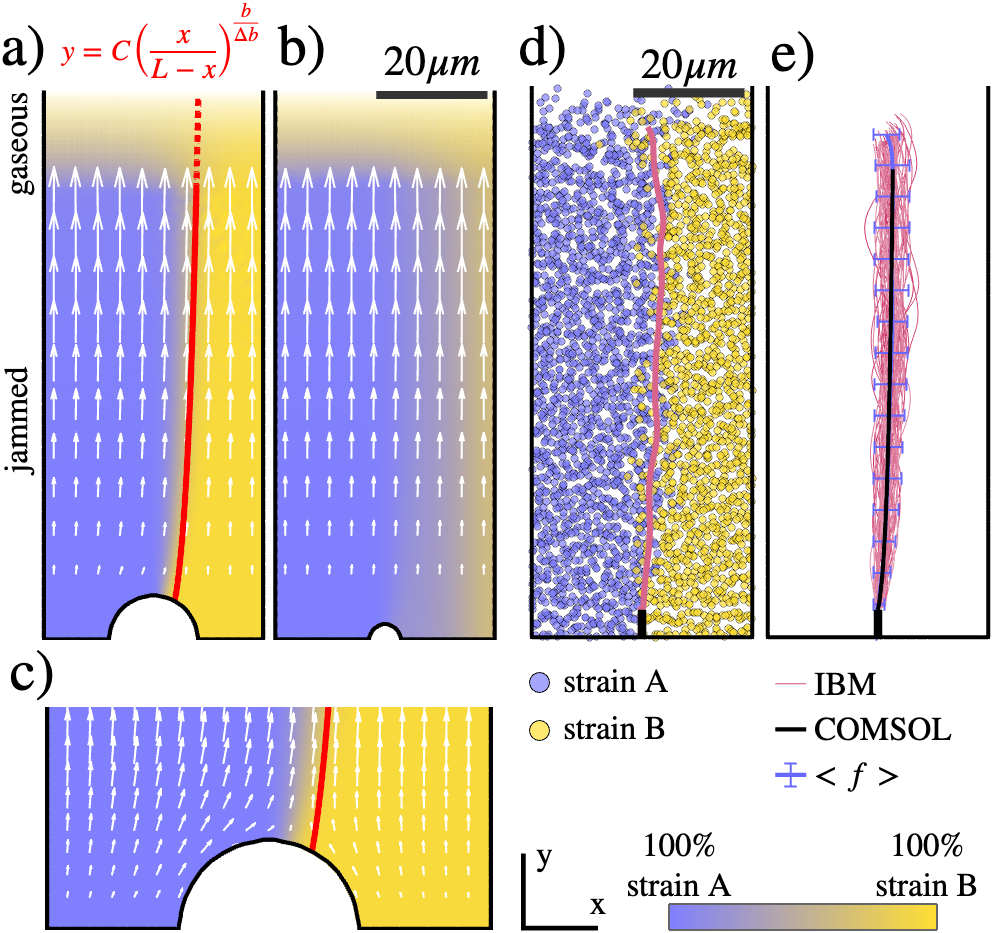}
\caption{Defect can enable two strain coexistence. a)-c) COMSOL simulation. a)  A round-shaped defect at the cavity floor can prevent fixation of the faster-growing strain (blue) if its size exceeds the critical threshold given by Eq.~\ref{eq:circular}. The stable boundary  shape is predicted by Eq.~\ref{eq:bound-shape} (red). 
(b) If the defect is smaller than the threshold, the faster-growing strain fixates as expected.
(c) Zoom-in on the velocity field (white arrows) near the cavity floor of (a). In cases where coexistence occurs, the slower-growing strain is protected by the velocity field being redirected along the defect, effectively shielding it. 
(d) The same effect is observed in individual-based model (IBM) simulations for a pin-shaped defect; an exemplary simulation is shown with the magenta line marking the instant boundary between two strains.
(e) Averaging over multiple IBM simulations (magenta) reveals that the resulting boundary (light blue) aligns closely with the analytical prediction (black curve), confirming the robustness of the theoretical result.}
\label{fig:boundary-shape}
\end{figure}

In contrast, geometry plays a crucial role for the population dynamics in the jammed state. In the previous section, we showed that a defect can protect a neutral strain from extinction caused by stochastic fluctuations. Now, we demonstrate that such a defect (see Fig. \ref{fig:boundary-shape}) can also enable the stable coexistence of two strains -- even when one has a significant fitness advantage.

In the jammed limit, we can solve Eq.~\ref{eq:react-diff} analytically and obtain the shape of the interstrain boundary (see SI Section~\ref{note:Note5}):
\begin{equation}
y_b(x) = C \left( \frac{x}{L - x} \right)^\frac{1}{\epsilon},
\label{eq:bound-shape}
\end{equation}
where $x$ is directed along the width of the cavity, $y$ along the length, and $L$ is the width of the cavity (see Fig.~\ref{fig:boundary-shape}A) . The parameter $\epsilon = \frac{\Delta g}{g_{\rm average}}$ quantifies the relative growth rate difference between the two strains, and the constant $C$ is determined by the shape of the defect at the cavity floor.  Notably, the average interstrain boundary observed in the individual-based model aligns well with this analytical prediction (see Fig. \ref{fig:boundary-shape}E).

If the defect is too small,  the faster-growing strain still will spread across the cavity, resulting in the loss of a stable boundary. The critical defect size required to sustain coexistence can be derived from the boundary shape, under the condition that the boundary between the two strains must be perpendicular to the cavity floor for smooth defects (see SI Section~\ref{note:Note5}).

For example, in the case of a circular defect centered in the middle of the cavity floor, stable coexistence occurs if the defect’s radius exceeds a critical threshold:
\begin{equation}
    r > r_{\rm cr} = \frac{L \epsilon}{2}
    \label{eq:circular}
\end{equation}

(see SI Section~\ref{note:Note5} for derivation). 

\subsection*{Friction and adhesion to the cavity wall can stabilize strain coexistence}
\label{sec:friction}

    \begin{figure}[h]
\includegraphics[width=\linewidth]{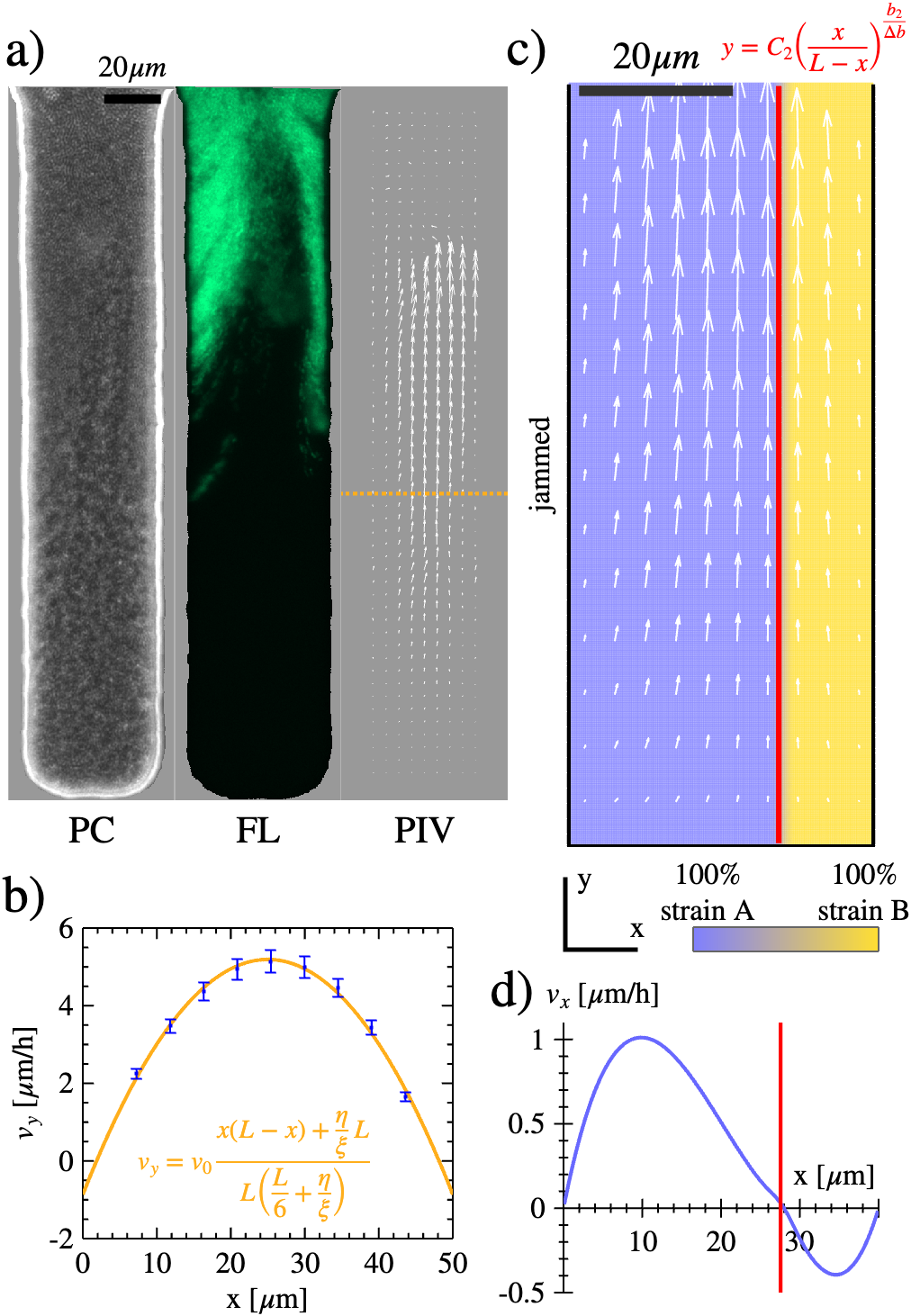}
\caption{Friction and adhesion to the cavity wall stabilize strain coexistence.
(a) Microscopy images of an exemplary cavity of a microfluidic device filled with a two strain population ({\it Acetobacter pasterianus} wild type and GFP). Left: phase-contrast image of the filled cavity. Middle: fluorescence image with a green fluorescing strain. Due to adhesion to the wall the cells are not pushed out of the cavity. Right: velocity field reconstructed from the time-lapse recording of this particular cavity, using PIVlab.
(b) For neutral strains, the Stokes equation predicts a parabolic velocity profile along the cavity (orange curve; see SI Eq.~\ref{eq:parabolic}). This prediction agrees well with experimental measurements (blue error bars) obtained from time-averaged PIV data near the center of the cavity (marked by the orange line in panel a).
(c) In the case of strains with different growth rates, COMSOL solution predicts a stable coexistence state in the presence of friction to the wall. The red line indicates the analytically predicted position of the interstrain boundary (Eq.~\ref{eq:friction-x}). White arrows represent the velocity field, which points slightly inwards the cavity. At the interstrain boundary, the horizontal ($x$) component of the velocity vanishes.
(d) $x$ component of velocity ($v_x$), confirming that it crosses zero at the predicted boundary position. }
\label{fig:friction}
\end{figure}

While the reaction-diffusion model Eq.~\ref{eq:react-diff} provides a quantitatively accurate description of the gaseous phase and qualitatively -- of the jammed phase~\cite{Karita2022}, it comes with inherent limitations. Notably, it cannot account for friction or adhesion to the cavity walls. Experimental observations (see Fig.~\ref{fig:friction}A and SI Section~\ref{Note:Experiment} for the experimental details) show that both friction and wall adhesion are frequently present and significantly influence cell motion within the cavity. Indeed, in Fig.~\ref{fig:friction}A one may see that cells of a green-fluorescent strain adhere to the cavity wall and thereby can not be pushed out of the cavity by the flow of the proliferating wild type cells, thus allowing stable coexistence between.

A standard approach for describing continuous media dynamics is the Navier-Stokes equation. In the regime of slow motion, where inertial effects can be neglected, the system simplifies to the Stokes equation. In this framework, the stress tensor $\sigma_{i j}$ for a fluid with viscosity $\eta$ is  given by:
\begin{equation}
   \sigma_{i j} = \eta (\partial_i v_j + \partial_j v_i) - p \delta_{i j}, 
   \label{eq:stress}
\end{equation}

subject to the conditions of force balance\footnote{Here, we neglect friction with the surrounding water, which does not alter the resulting velocity field — see SI Section~\ref{note:Note6} for details.} and incompressibility:

\begin{equation}
    \partial_i \sigma_{i j} = 0 \mbox{ and }\partial_i v_{i (\alpha) } = g_{(\alpha)}
    \label{eq:Stokes}
\end{equation}

 The incompressibility condition differs from that in the standard Navier-Stokes equation by incorporating the growth rate of the cells. This approach has been previously implemented, for example, by \cite{Giometto2018}.

Friction to the walls imposes boundary conditions on the stress tensor. Specifically, the shear stress at the wall is proportional to the velocity along the wall:
\begin{equation}
    \sigma_{x y}( 0) - \xi v_y( 0)  = 0
\end{equation}

\begin{equation}
    \sigma_{x y}( L) + \xi v_y(  L)  = 0
\end{equation}

Friction alters the velocity field in two principal ways. First, it reduces the flow velocity near the cavity walls relative to the center, as shown in Fig.~\ref{fig:friction}B, which compares the $y$-component of the predicted velocity field with experimental measurements obtained via particle image velocimetry (PIV) \cite{thielicke2022pulse, thielicke2021particle}. Second, friction generates an additional inward flux, directing cells away from the cavity walls towards the central axis of the cavity.


The origin of this flux becomes evident in the extreme case, where friction is so strong that cells adhere to the wall and are unable to move along it. In this scenario, cell proliferation can only occur in the direction perpendicular to the wall, pushing new cells toward the center of the cavity. This condition is expressed as $\partial_x v_x (0) = \partial_x v_x (L) =  g$.

This inward flow creates a protected position for the cells near the wall. Thus, even when a faster growing strain occupies the opposite wall, the inward flow due do the friction stops the motion of interstrain boundary due to the growth rate difference (see Fig.~\ref{fig:friction}C,D).

For a smooth floor (i.e., no obstacles), the position of the interstrain boundary is determined by the relative growth rate difference and the dimensionless friction parameter $ q =  \frac{L \xi}{\eta} $, leading to the expression:

\begin{equation}
    x_b = L \frac{1}{\epsilon} \left( \sqrt{1  + \left( 2 + \frac{6}{q} \right)  \epsilon^2} - 1 \right) + L /2  \approx  L \epsilon ( 1 + \frac{3}{q} ) + L / 2
    \label{eq:friction-x}
\end{equation}

(see derivation on the SI Section~\ref{note:Note7}).

\section*{Conclusions}
Our findings demonstrate that the stable coexistence of two microbial strains within the same cavity is possible, even when one strain has a growth advantage. This challenges the intuition derived from well-mixed systems, where the fitter strain inevitably outcompetes and completely replaces the slower-growing one. Our analysis reveals that this effect does not require a physical barrier between the strains. Instead, a defect significantly smaller than the cavity length can lead to a stable boundary between two populations spanning throughout the entire cavity. Similarly, friction and adhesion to the walls may have the same effect. In both cases, the cells at the boundary remain in direct contact with each other, yet the slower-growing strain can persist.

\begin{figure}
\centering
\includegraphics[width=1\linewidth]{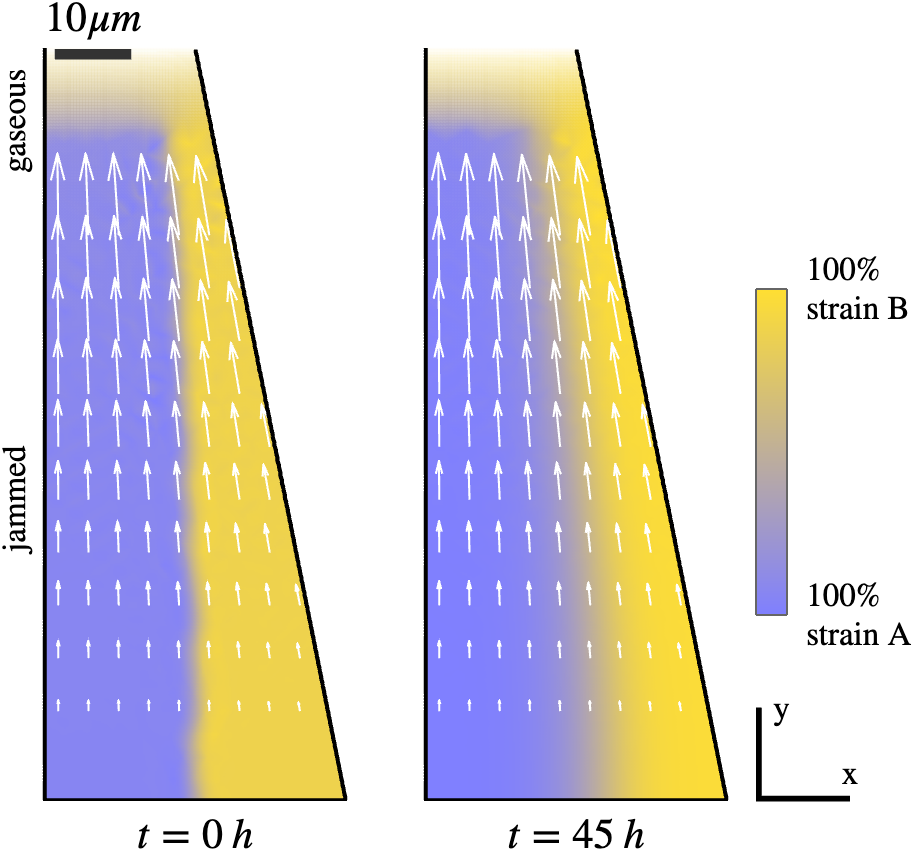}
\caption{The tilted wall stabilizes coexistence of two strains. Left panel:  Initial state with strain A (blue) with growth rate $g_A = 0.33 h^{-1} $ and strain B (yellow) with  a growth rate $g_B = 0.22 h^{-1} $. Right panel: Stable coexistence of two strains. Velocity field (white arrows) directed outwards of the region populated by the slower-growing strain B. }
\label{fig:prism}
\end{figure} 
The key mechanism behind this stable coexistence lies in the velocity profile of the system. Stability arises when a region exists in which the velocity field is directed outward of this region (and usually inward to the cavity), either due to the geometry of the cavity or interactions with its boundaries. If this region is occupied by a slower growing strain, it is harder for a faster growing strain to invade the region and push the slower grain strain out. One scenario is the cavity itself, which is extremely protected from the invasion outside when it is in the jammed state.  Here we also found the same effect for the defect at the floor of the cavity, where the curved velocity profile near the defect generates an outward flow, not allowing faster growing strain to get around the defect.  Similarly, in case of the friction along the walls the velocity profile ensures that region near the wall has an outward flow toward the middle of the cavity, not letting the faster growing strain to get closer to the wall.

These scenarios do not represent all possible ways in which such a stabilizing velocity field can emerge. For example, as shown in Fig.~\ref{fig:prism}, a cavity with a tilted wall can generate a velocity flow directed from the tilted wall to the central part of the cavity. If the slower-growing strain initially occupies the corner adjacent to the tilted wall, the outward-directed flow from this region helps maintain its presence, preventing complete displacement by the faster-growing strain.

While microfluidic devices provide a controlled environment for studying these effects, natural ecosystems are far more complex. Nevertheless, it is plausible that regions with outward-directed velocity fields are common in nature, arising from diverse factors such as surface interactions, geometric constraints, and hydrodynamic flows. This suggests that strain or even species coexistence could occur without the need for physical barriers. Such a mechanism may be particularly relevant in ecological systems where community stability depends on species interactions, such as cross-feeding. Unlike physical barriers, which can restrict nutrient diffusion, boundaries induced by velocity field allow for nutrient exchange, supporting the persistence of diverse microbial populations.

\begin{acknowledgements}
This work was supported by the Deutsche Forschungsgemeinschaft (DFG) within the SPP 2389 priority program “Emergent Functions of Bacterial Multicellularity” (Grant number: 503995533). OH acknowledges support by a Humboldt Professorship of the Alexander von Humboldt Foundation. VPY would like to thank the "Fundació La Caixa" for support through its Postgraduate Fellowship Programme. The project that gave rise to these results received the support of a fellowship from the ”la Caixa” Foundation (ID 100010434). The fellowship code is LCF/BQ/EU21/11890104. We used ChatGPT-4o to refine the style of our text. VS and CW would like to thank Katja Taute for helpful comments and suggestions. 
\end{acknowledgements}

\section*{Bibliography}
\bibliography{references}

\begin{thebibliography}{32}
\providecommand{\natexlab}[1]{#1}
\providecommand{\url}[1]{\texttt{#1}}
\expandafter\ifx\csname urlstyle\endcsname\relax
  \providecommand{\doi}[1]{doi: #1}\else
  \providecommand{\doi}{doi: \begingroup \urlstyle{rm}\Url}\fi

\bibitem[Karita et~al.(2022)Karita, Limmer, and Hallatschek]{Karita2022}
Yuya Karita, David~T. Limmer, and Oskar Hallatschek.
\newblock Scale-dependent tipping points of bacterial colonization resistance.
\newblock \emph{Proceedings of the National Academy of Sciences}, 119\penalty0 (7):\penalty0 e2115496119, February 2022.
\newblock \doi{10.1073/pnas.2115496119}.
\newblock Edited by David Weitz.

\bibitem[Postek et~al.(2024)Postek, Staśkiewicz, Lilja, and Wacław]{postek24}
Witold Postek, Klaudia Staśkiewicz, Elin Lilja, and Bartłomiej Wacław.
\newblock Substrate geometry affects population dynamics in a bacterial biofilm.
\newblock \emph{Proceedings of the National Academy of Sciences}, 121\penalty0 (17):\penalty0 e2315361121, 2024.
\newblock \doi{10.1073/pnas.2315361121}.

\bibitem[Gause(1934)]{Gause1934}
Georgii~Frantsevich Gause.
\newblock \emph{The Struggle For Existence}.
\newblock Williams \& Wilkins, Baltimore, 1 edition, 1934.

\bibitem[Hardin(1960)]{hardin1960competitive}
Garrett Hardin.
\newblock The competitive exclusion principle.
\newblock \emph{Science}, 131\penalty0 (3409):\penalty0 1292--1297, 1960.
\newblock \doi{10.1126/science.131.3409.1292}.

\bibitem[Conwill et~al.(2022)Conwill, Kuan, Damerla, Poret, Baker, Tripp, Alm, and Lieberman]{conwill2022anatomy}
Arolyn Conwill, Anne~C Kuan, Ravalika Damerla, Alexandra~J Poret, Jacob~S Baker, A~Delphine Tripp, Eric~J Alm, and Tami~D Lieberman.
\newblock Anatomy promotes neutral coexistence of strains in the human skin microbiome.
\newblock \emph{Cell Host \& Microbe}, 30\penalty0 (2):\penalty0 171--182, 2022.

\bibitem[Lee et~al.(2013)Lee, Donaldson, Mikulski, and et~al.]{lee2013bacterial}
S.~M. Lee, N.~E. Donaldson, A.~Mikulski, and et~al.
\newblock Bacterial colonization factors control specificity and stability of the gut microbiota.
\newblock \emph{Nature}, 501:\penalty0 426--429, 2013.
\newblock \doi{10.1038/nature12447}.

\bibitem[Martínez and et~al.(2018)]{martinez2018experimental}
I.~Martínez and et~al.
\newblock Experimental evaluation of the importance of colonization history in early-life gut microbiota assembly.
\newblock \emph{eLife}, 7:\penalty0 e36521, 2018.
\newblock \doi{10.7554/eLife.36521}.

\bibitem[Obadia and et~al.(2017)]{obadia2017probabilistic}
B.~Obadia and et~al.
\newblock Probabilistic invasion underlies natural gut microbiome stability.
\newblock \emph{Current Biology}, 27:\penalty0 1999--2006.e8, 2017.
\newblock \doi{10.1016/j.cub.2017.05.034}.

\bibitem[Lawley and Walker(2013)]{lawley2013intestinal}
T.~D. Lawley and A.~W. Walker.
\newblock Intestinal colonization resistance.
\newblock \emph{Immunology}, 138:\penalty0 1--11, 2013.
\newblock \doi{10.1111/imm.12003}.

\bibitem[Donaldson et~al.(2016)Donaldson, Lee, and Mazmanian]{donaldson2016gut}
G.~P. Donaldson, S.~M. Lee, and S.~K. Mazmanian.
\newblock Gut biogeography of the bacterial microbiota.
\newblock \emph{Nature Reviews Microbiology}, 14\penalty0 (1):\penalty0 20--32, 2016.
\newblock \doi{10.1038/nrmicro3552}.

\bibitem[MacArthur and Wilson(1967)]{macarthur1967island}
Robert~H. MacArthur and Edward~O. Wilson.
\newblock \emph{The Theory of Island Biogeography}.
\newblock Princeton University Press, 1967.
\newblock ISBN 9780691088365.

\bibitem[Wu and Vankat(1995)]{wu1995island}
Jianguo Wu and John~L. Vankat.
\newblock Island biogeography: Theory and applications.
\newblock In \emph{Encyclopedia of Environmental Biology}, volume~2, pages 371--379. 1995.

\bibitem[Seth and Taga(2014)]{seth2014nutrient}
Erica~C. Seth and Michiko Taga.
\newblock Nutrient cross-feeding in the microbial world.
\newblock \emph{Frontiers in Microbiology}, 5:\penalty0 350, 2014.
\newblock \doi{10.3389/fmicb.2014.00350}.
\newblock Review.

\bibitem[Smith et~al.(2019)Smith, Shorten, Altermann, Roy, and McNabb]{smith2019classification}
Nick~W. Smith, Paul~R. Shorten, Eric Altermann, Nicole~C. Roy, and Warren~C. McNabb.
\newblock The classification and evolution of bacterial cross-feeding.
\newblock \emph{Frontiers in Ecology and Evolution}, 7:\penalty0 153, 2019.
\newblock \doi{10.3389/fevo.2019.00153}.
\newblock Review article, Section: Population, Community, and Ecosystem Dynamics.

\bibitem[Mataigne et~al.(2021)Mataigne, Vannier, Vandenkoornhuyse, and Hacquard]{mataigne2021microbial}
Victor Mataigne, Nathan Vannier, Philippe Vandenkoornhuyse, and Stéphane Hacquard.
\newblock Microbial systems ecology to understand cross-feeding in microbiomes.
\newblock \emph{Frontiers in Microbiology}, 12:\penalty0 780469, 2021.
\newblock \doi{10.3389/fmicb.2021.780469}.
\newblock Mini Review article, Section: Microbial Symbioses.

\bibitem[Flint et~al.(2012)Flint, Scott, Duncan, Louis, and Forano]{flint2012}
Harry~J. Flint, Karen~P. Scott, Sylvia~H. Duncan, Petra Louis, and Evelyne Forano.
\newblock Microbial degradation of complex carbohydrates in the gut.
\newblock \emph{Gut Microbes}, 3:\penalty0 289--306, 2012.

\bibitem[LeBlanc et~al.(2013)LeBlanc, Milani, de~Giori, Sesma, Van~Sinderen, and Ventura]{leblanc2013}
Jean~Guy LeBlanc, Christian Milani, Graciela~S. de~Giori, Fernando Sesma, Douwe Van~Sinderen, and Marco Ventura.
\newblock B-group vitamin production by lactic acid bacteria–current knowledge and potential applications.
\newblock \emph{Journal of Applied Microbiology}, 115:\penalty0 1297--1309, 2013.

\bibitem[Bacher et~al.(2000)Bacher, Eberhardt, Fischer, Kis, and Richter]{bacher2000}
A~Bacher, S~Eberhardt, M~Fischer, K~Kis, and G~Richter.
\newblock Biosynthesis of vitamin b2 (riboflavin).
\newblock \emph{Annual Review of Nutrition}, 20:\penalty0 153--167, 2000.
\newblock \doi{10.1146/annurev.nutr.20.1.153}.

\bibitem[Rakoff-Nahoum et~al.(2005)Rakoff-Nahoum, Paglino, Eslami-Varzaneh, Edberg, and Medzhitov]{rakoff2005}
Seth Rakoff-Nahoum, Justin Paglino, Feroze Eslami-Varzaneh, Sara Edberg, and Ruslan Medzhitov.
\newblock Recognition of commensal microflora by toll-like receptors is required for intestinal homeostasis.
\newblock \emph{Cell}, 118:\penalty0 229--241, 2005.

\bibitem[Louis et~al.(2014)Louis, Hold, and Flint]{louis2014}
Petra Louis, Georgina~L. Hold, and Harry~J. Flint.
\newblock The gut microbiota, bacterial metabolites and colorectal cancer.
\newblock \emph{Nature Reviews Microbiology}, 12:\penalty0 661--672, 2014.

\bibitem[Batchelor(1976)]{Batchelor_1976}
G.~K. Batchelor.
\newblock Brownian diffusion of particles with hydrodynamic interaction.
\newblock \emph{Journal of Fluid Mechanics}, 74\penalty0 (1):\penalty0 1–29, 1976.
\newblock \doi{10.1017/S0022112076001663}.

\bibitem[Schreck et~al.(2023)Schreck, Fusco, Karita, Martis, Kayser, Duvernoy, and Hallatschek]{schreck2023impact}
C.~F. Schreck, D.~Fusco, Y.~Karita, S.~Martis, J.~Kayser, M.~C. Duvernoy, and O.~Hallatschek.
\newblock Impact of crowding on the diversity of expanding populations.
\newblock \emph{Proceedings of the National Academy of Sciences of the United States of America}, 120\penalty0 (11):\penalty0 e2208361120, 2023.
\newblock \doi{10.1073/pnas.2208361120}.
\newblock Epub 2023 Mar 7.

\bibitem[Hallatschek and Nelson(2008)]{HALLATSCHEK2008158}
Oskar Hallatschek and David~R. Nelson.
\newblock Gene surfing in expanding populations.
\newblock \emph{Theoretical Population Biology}, 73\penalty0 (1):\penalty0 158--170, 2008.
\newblock ISSN 0040-5809.
\newblock \doi{https://doi.org/10.1016/j.tpb.2007.08.008}.

\bibitem[Moran(1962)]{Moran1962}
P.~A.~P. Moran.
\newblock \emph{Statistical Processes of Evolutionary Theory}.
\newblock Oxford University Press, Oxford, 12 1962.
\newblock ISBN 9780198543435.

\bibitem[Giometto et~al.(2018)Giometto, Nelson, and Murray]{Giometto2018}
Andrea Giometto, David~R. Nelson, and Andrew~W. Murray.
\newblock Physical interactions reduce the power of natural selection in growing yeast colonies.
\newblock \emph{Proceedings of the National Academy of Sciences}, 115\penalty0 (45):\penalty0 11448--11453, 10 2018.
\newblock \doi{10.1073/pnas.1809587115}.
\newblock ORCID: Giometto: 0000-0002-0544-6023, Murray: 0000-0002-0868-6604.

\bibitem[Thielicke(2022)]{thielicke2022pulse}
W.~Thielicke.
\newblock Pulse-length induced motion blur in {PIV} particle images: {To} be avoided at any cost?
\newblock In \emph{Proceedings of the {Fachtagung Experimentelle Strömungsmechanik} 2022}, Ilmenau, Germany, September 2022.
\newblock ISBN 978-3-9816764-8-8.

\bibitem[Thielicke and Sonntag(2021)]{thielicke2021particle}
William Thielicke and Ren\'{e} Sonntag.
\newblock Particle image velocimetry for {MATLAB}: {Accuracy} and {Enhanced} {Algorithms} in {PIVlab}.
\newblock \emph{Journal of Open Research Software}, 9\penalty0 (1):\penalty0 12, May 2021.
\newblock \doi{10.5334/jors.334}.

\bibitem[Wang and et~al.(2010)]{wang2010robust}
P.~Wang and et~al.
\newblock Robust growth of \textit{Escherichia coli}.
\newblock \emph{Current Biology}, 20:\penalty0 1099--1103, 2010.
\newblock \doi{10.1016/j.cub.2010.04.045}.

\bibitem[Lambert and Kussell(2014)]{lambert2014memory}
G.~Lambert and E.~Kussell.
\newblock Memory and fitness optimization of bacteria under fluctuating environments.
\newblock \emph{PLoS Genetics}, 10\penalty0 (9):\penalty0 e1004556, 2014.
\newblock \doi{10.1371/journal.pgen.1004556}.

\bibitem[Yamada et~al.(2015)Yamada, Deshpande, Bruce, Mak, and Ja]{yamada2015microbes}
R.~Yamada, S.~A. Deshpande, K.~D. Bruce, E.~M. Mak, and W.~W. Ja.
\newblock Microbes promote amino acid harvest to rescue undernutrition in \textit{Drosophila}.
\newblock \emph{Cell Reports}, 10:\penalty0 865--872, 2015.
\newblock \doi{10.1016/j.celrep.2015.01.018}.

\bibitem[Darcy(1856)]{Darcy1856}
Henry Darcy.
\newblock \emph{Les fontaines publiques de la ville de Dijon}.
\newblock Dalmont, Paris, 1856.

\bibitem[Whitaker(1986)]{Whitaker1986}
S.~Whitaker.
\newblock Flow in porous media {I}: {A} theoretical derivation of {Darcy's} law.
\newblock \emph{Transport in Porous Media}, 1:\penalty0 3--25, 1986.
\newblock \doi{10.1007/BF01036523}.

\end{thebibliography}

\onecolumn
\newpage

\captionsetup*{format=largeformat}
\section{Experiment setup}
\label{Note:Experiment}

The experimental design follows the approach described in~\cite{Karita2022}, using a microfluidic incubation platform specifically built to track bacterial population dynamics over extended periods across a range of spatial scales (Fig.~\ref{fig:intro-fig}). The device features a main supply channel that continuously delivers fresh growth media, allowing for stable environmental conditions to be maintained for several days. As bacterial cells are introduced into the system and travel through the main channel, they encounter a series of rectangular side chambers (or "crypts") with depths ranging from 10 to 350 $\mu$m. Although fluid flow inside these side cavities is minimal, nutrient availability remains high due to the efficient diffusion of small molecules from the supply channel\cite{wang2010robust, lambert2014memory}, enabling sustained bacterial growth.\\
By examining colonization behavior across chambers of varying depths, we can identify how ecological processes depend on spatial scale. To facilitate such comparisons within a single microscopy field of view, the chambers were arranged in order of increasing depth. This configuration gives the device a stepped appearance, reminiscent of a pan flute, which led us to refer to it as the "microfluidic pan flute". \\
In this study, we use the platform to investigate colonization dynamics primarily in Acetobacter, an aerobic genus commonly found in the Drosophila gut~\cite{obadia2017probabilistic, yamada2015microbes}, although the design is suitable for a range of bacterial taxa. \\
Microfluidic channels were produced, as described previously~\cite{Karita2022}. \textit{Acetobacter pasteurianus} wildtype and strain ZTG272 (labeled with GFP) were grown in MRS medium, and mixed 1:1 before being loaded into a microfluidic channel. Loading and later MRS medium supply was facilitated through a 5~ml syringe (Hamilton Bonaduz) and connected PTFE tubing. The syringe was mounted onto a syringe pump (Cetoni), which created a constant flow at a flow rate of 20 $\mu$l/h. The fully prepared microfluidic chip was mounted onto a Zeiss AxioObserver.Z1 microscope equipped with a climate chamber (Incubation system S, Pecon, set to 30$^o$C throughout the recording). Bacterial growth was imaged through a 20x LD Plan-NEOFLUAR phase contrast objective (Zeiss, NA=0.4), GFP fluorescence was excited at 488 nm (Zeiss Colibri2) and the images were recorded by a Zeiss Axiocam705.

\section{Comparing flow velocity in the microfluidic device with the COMSOL simulation} \label{note:Note1}

We use COMSOL Multiphysics 6.2 to numerically solve the Stokes equation for water flow through the microfluidic device. The device is designed in AutoCAD 2024, and the resulting file is exported to COMSOL to construct the simulation geometry. Starting from a two-dimensional .dxf file generated in AutoCAD, we create a three-dimensional geometry using the Extrusion function, with device thickness set to 20, 30, or 40 $\mu$m depending on the simulation.

Simulation results are exported and analyzed using custom Python scripts (see code repository on \href{https://github.com/valentinslepukhin/self-organised-seclusion}{GitHub} ). We observe that the velocity magnitude within the cavity decreases exponentially with distance from the opening, following the relation $v \propto e^{-k y}$ where the decay constant $k$ decreases as the cavity width increases. Specifically, for a cavity width of 50\,$\mu$m, we find $k \approx 0.17\,\mu\text{m}^{-1}$, while for a width of 80\,$\mu$m, $k \approx 0.123\,\mu\text{m}^{-1}$.

This slower decay in wider cavities results in a significantly smaller effective cavity size. Taking a threshold velocity of 0.3\,$\mu$m/s (below which cells are not flushed away), as used in~\cite{Karita2022}, we estimate a penetration depth of approximately 75--95\,$\mu$m for a 50\,$\mu$m wide cavity, and 130--180\,$\mu$m for an 80\,$\mu$m wide cavity. Changes in flow rate have only a marginal effect: for example, reducing the flow rate by a factor of two increases the effective cavity length by only $\ln(2)/k$, which corresponds to just a few microns. Therefore, to achieve improved flow control, it is advisable to either keep the cavities narrow or increase their depth substantially.

\section{Individual-based model} 
\label{note:Note2}

In the individual-based model, microbes are treated as point-like particles subjected to Brownian motion.

The position $\bm{x}_i$ of microbe $i$ evolves according to an overdamped Langevin equation, written in Itō discretization as:
\begin{equation}\label{eq:2d_langevin}
    \dot{\bm{x}}_i = \sqrt{2 D_{\rm self}(\rho)} \cdot \bm{\zeta_i}(t) + \bm{v}_{\rm drift},
\end{equation}
where where $\zeta(t)$ is unitary, uncorrelated Gaussian white noise satisfying $\langle \zeta_i(t)\zeta_i(t+\Delta t)\rangle = \delta_{\Delta t,0}$ and $\langle \zeta_i(t)\zeta_j(t) = \delta_{i,j} \rangle$, and $\bm{v}_{\rm drift}$ is the drift velocity given by:
\begin{equation}\label{eq:2d_v_drift}
    \bm{v}_{\rm drift} = \left(D_{\rm self}(\rho) - D_{\rm col}(\rho)\right) \bm{\nabla}_{\bm{x}_i} \ln \rho + \bm{\nabla}_{\bm{x}_i} D_{\rm self}(\rho).
\end{equation}

Here, $D_{\rm self}(\rho)$ and $D_{\rm col}(\rho)$ are the self-diffusivity and collective diffusivity, respectively, both of which depend on the local cell density $\rho(\bm{x}_i,t)$. For brevity, we will often omit the explicit position and time dependence of $\rho$.

The self-diffusivity $D_{\rm self}(\rho)$ determines the magnitude of individual Brownian displacements, while the collective diffusivity $D_{\rm col}(\rho)$ governs the macroscopic fluxes of microbes, as described in Equation~\ref{eqSI:FokkerPlanck_without_births}. The expression for $D_{self}$ is:

\begin{equation}
    D_{\rm self} = D_0 \cdot
    \begin{cases}
         (1-0.95 \cdot \rho/\rho_{\rm jam}), \quad \rho<\rho_{\rm jam}\\
        1.5\times 10^{-3} , \quad \rho \ge \rho_{\rm jam}
    \end{cases}
\end{equation}

where $\rho_{\rm jam}$ is the cell density at the jamming transition, given by the packing fraction $\Phi(\rho) =\rho/A_{\rm  cell}$ and $A_{\rm cell} = \pi (d_{\rm cell}/2)^2$ the area of a cell of diameter $d_{\rm cell}$, at the jamming transition. The value $\Phi(\rho_{\rm jam}) \equiv \Phi_{\rm jam} =0.64$ is taken from the jamming transition of hard spheres, as in \cite{Karita2022}. The expression for $D_{\rm col}$ is given by:

\begin{equation}
D_{\rm col}(\rho)=D_0 \cdot
\begin{cases}
     \frac{\mu(\Phi(\rho)) \cdot dP(\Phi(\rho))}{\mu(0)dP(0)}, \quad \rho \le\rho_{\rm  jam} \\
      1000 \cdot (\rho-\rho_{\rm  jam}) /D_{col}(\rho_{\rm  jam}),  \quad\rho>\rho_{\rm  jam}  
\end{cases}
\end{equation}

where $dP = 6/\pi\cdot A(\Phi) + 6 \Phi/\pi \cdot \partial_\Phi A(\Phi)$, $A(\Phi) = (1+\Phi +\Phi^2 -\Phi^3)/(1-\Phi)^3$, as in \cite{Karita2022}.

In addition to stochastic motion, each microbe reproduces stochastically with a growth rate $g_i$. Cells that reach the open end of the cavity are removed from the simulation. If a microbe's trajectory intersects a cavity wall it undergoes an elastic collision and its position is mirrored with respect to the wall at the next time step.

To estimate the density field $\rho(\bm{x}_i,t)$, the cavity space is discretized into a square grid with spacing $\Delta = 0.5$. Density is measured in each grid square and then smoothed using a low-pass Gaussian filter with window size $s = 5$ and grid spacing $\Delta = 0.5$. Gradients of $\rho$ and other quantities are computed on the grid using finite difference methods. For cells located closer than $\Delta$ to the wall, the drift velocity is assumed to be $v_{\rm drift} = 0$.


To demonstrate that the Langevin description used in the individual-based model is consistent with our continuous reaction-diffusion framework, we begin by considering the general form of the Fokker–Planck equation corresponding to an Itō-discretized stochastic process:

\begin{equation}\label{eq:multid_generic_FP}
\begin{split}
    \partial_t p(\bm{x},t) &= - \bm{\nabla}_{\bm{x}} \cdot \left[ 
    \bm{F} (\bm{x})\, p(\bm{x},t) 
    \right]
    + \bm{\nabla}_{\bm{x}} \cdot \bm{\nabla}_{\bm{x}} \cdot \left[
    \bm{G}(\bm{x})\, p(\bm{x},t) 
    \right] \\
    &= - \sum_i  \partial_{x_i}  \left[
    F_{i} (\bm{x})\, p(\bm{x},t) \right]
    + \sum_{i,j} \partial_{x_i} \partial_{x_j} \left[
    G_{ij}(\bm{x})\, p(\bm{x},t) 
    \right],
\end{split}
\end{equation}

where the corresponding Langevin equation is given by:

\begin{equation}
    \dot{\bm{x}} = \bm{F}(\bm{x}) + \bm{G}(\bm{x}) \cdot \bm{\zeta}(t),
\end{equation}

with $\bm{\zeta}(t)$ denoting Gaussian white noise.

In our model, we assume uncorrelated noise, i.e., $G_{ij} = 0$ for $i \ne j$, and set $G_{ii} = 2 D_{\rm self}(\rho)$. Identifying the drift term as $\bm{F}(\bm{x}) \equiv \bm{v}_{\rm drift}$ and substituting into Equation~\ref{eq:multid_generic_FP}, we obtain the following Fokker–Planck equation:

\begin{equation}\label{eqSI:FokkerPlanck_without_births}
    \partial_t \rho = \bm{\nabla} \cdot \left( D_{\rm col}(\rho)\, \bm{\nabla} \rho \right),
\end{equation}

which matches Equation~\ref{eq:react-diff} in the main text for a single strain, up to the omission of the birth term.

\begin{figure}
\centering
\includegraphics[width=.5\linewidth]{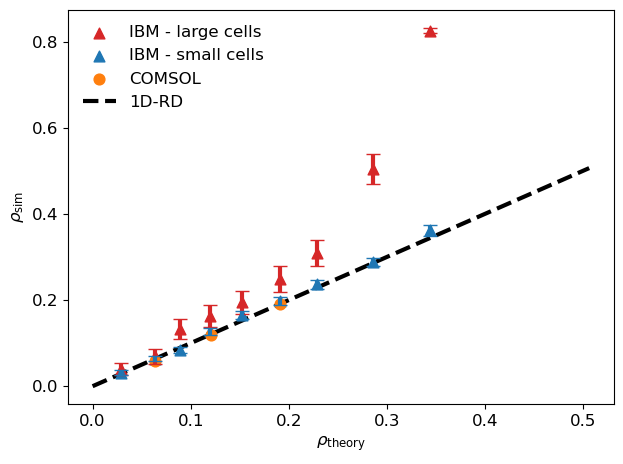}
\caption{Comparison between the individual-based model (triangular markers with error bars), the continuous model (COMSOL, orange circular markers) and the expectation from the 1D reaction diffusion system from \cite{Karita2022} (black dashed line) for the cell density at the  cavity bottom in steady state. The graph shous analytical prediction vs results of the numerical simulations. In the individual-based simulations, the number of cells in the cavity can be controlled by setting their nominal diameter, $d$. Blue markers: For small cells ($d=1/\sqrt{10}$), the system is close to the continuous ($N \sim10^3-10^4$ cells)  and the density profile matches the expectation from the continuous model. Red markers: For large cells ($d=1$), finite population effects ($N \sim 10^2-10^3$) cause an increase in cell density with respect to the continuous predictions, promoting jamming in otherwise gaseous cavities. Simulations shown here were ran with $D_0=1$ and $b \sim 10^{-4}$ to highlight finite-size effects.}
    \label{fig:copmarison_with_comsol}
\end{figure}

In the large-$N$ limit, this macroscopic description converges to the continuous model. However, for finite $N$, stochastic effects arising from the random birth process lead to deviations from the deterministic model (see Fig.~\ref{fig:copmarison_with_comsol}), including a noticeable increase in the average packing fraction within the cavity.

The code for the model is published on \href{https://github.com/valentinslepukhin/self-organised-seclusion}{GitHub}.

\section{Well-mixed case. Logistic growth} 
\label{note:Note3}

In the well-mixed limit, we assume that the relative frequency of each strain is spatially uniform. Thus, the density of strain $\alpha$ can be expressed as:

\begin{equation}
    \rho_\alpha(x,t) = c_\alpha(t) \rho(x,t),
\end{equation}

where $c_\alpha(t)$ is the time-dependent fraction of strain $\alpha$, and $\sum_\alpha c_\alpha = 1$.

Substituting this expression into the equation for the current of strain $\alpha$, we obtain:

\begin{equation}
    j_\alpha = - c_\alpha(t)\, D_{\rm col}(\rho)\, \nabla \rho.
\end{equation}

Using this in the continuity equation yields:

\begin{equation}
    \frac{\partial}{\partial t} (c_\alpha \rho) = c_\alpha\, \left[ \nabla \cdot (D_{\rm col}(\rho)\, \nabla \rho) + g_\alpha \rho \right].
\end{equation}

Expanding the time derivative on the left-hand side and simplifying gives:

\begin{equation}
    \frac{\partial c_\alpha}{\partial t} = \frac{c_\alpha}{\rho} \left[ \nabla \cdot (D_{\rm col}(\rho)\, \nabla \rho) + g_\alpha \rho - \frac{\partial \rho}{\partial t} \right].
\end{equation}

Now, consider the case of two strains, for which $c_A + c_B = 1$, implying $\dot{c}_A + \dot{c}_B = 0$. Substituting this condition into the above equation, we find:

\begin{equation}
    (g_A - g_B) c_A + g_B = - \frac{1}{\rho} \left[ \nabla \cdot (D_{\rm col}(\rho)\, \nabla \rho) - \frac{\partial \rho}{\partial t} \right].
\end{equation}

Inserting this back into the expression for $\dot{c}_1$, we obtain the following closed-form equation:

\begin{equation}
    \dot{c}_A = (g_A- g_B)\, c_A (1 - c_A),
\end{equation}

which is the classical logistic equation, with the corresponding analytical solution given in the main text.

\section{Jammed and incompressible. Velocity field and logistic growth} 
\label{note:Note4}

In the jammed phase, the total cell density remains close to $\rho_{\rm jam}$ throughout the cavity.\footnote{This assumption does not hold if the cells are rod-shaped and in a disordered phase, in which case additional considerations are required.} Since the self-diffusivity $D_{\rm self}$ is nearly zero in this regime, different strains do not mix. As a result, the strain density $\rho_\alpha$ at any given point is either $0$ or $\rho_{\rm jam}$.

When $\rho_\alpha = 0$, the corresponding cell current vanishes, i.e., $\vec{j}_\alpha = 0$. In regions where $\rho_\alpha = \rho_{\rm jam}$, we have:

\begin{eqnarray}
    \vec{j}_\alpha = D_{\rm col} \nabla \rho_\alpha
    \\
     \frac{\partial \rho_\alpha}{\partial t} + \nabla \cdot \vec{j}_\alpha  = g_\alpha \rho_\alpha
\end{eqnarray}

Note that even though the total density $\rho$ remains close to the jamming density $\rho_{\rm jam}$ everywhere in the jammed state, the term $D_{\rm col} \nabla \rho_\alpha$ is not necessarily small. This is because the collective diffusivity $D_{\rm col}$ diverges as the system approaches the jamming transition, which can compensate for the smallness of the density gradient.

To describe the total dynamics, we define a spatially varying growth rate field $g(x, y, t)$, such that $g(x, y, t) = g_\alpha$ in the region currently occupied by strain $\alpha$. With this, the total current and density evolve according to:

\begin{align}
    \vec{j} &= D_{\rm col} \nabla \rho, \\
    \frac{\partial \rho}{\partial t} + \nabla \cdot \vec{j} &= g(x, y, t)\, \rho.
\end{align}

Following~\cite{Karita2022}, we define the potential
\[
\Pi(\phi(x)) = \int_0^{\phi(x)} D_{\rm col}(\rho)\, d\rho,
\]
which allows us to rewrite the collective current as:

\begin{align}
    \vec{j} &= \nabla \Pi, \\
    \frac{\partial \rho}{\partial t} + \nabla \cdot \vec{j} &= g\, \rho.
\end{align}

In the jammed regime, the density is approximately uniform and close to the jamming value, i.e., $\rho \approx \rho_{\rm jam}$ everywhere. Under this assumption, we can simplify the equations:

\begin{align}
    \vec{j} &= \nabla \Pi, \\
    \nabla \cdot \vec{j} &= g\, \rho_{\rm jam}.
    \label{eq:divj_jammed}
\end{align}

It is convenient to introduce an effective velocity field defined by $\vec{v} = \vec{j} / \rho$, and a pressure field $p = \Pi / \rho_{\rm jam}$. In the jammed limit, where $\rho \approx \rho_{\rm jam}$, we obtain:

\begin{align}
    \vec{v} &= \nabla p, \\
    \nabla \cdot \vec{v} &= g.
\end{align}

We observe numerically that if the system is initialized with a vertical boundary separating two strains, the boundary remains vertical over time. This implies that the growth rate field can be written as:

\begin{equation}
    g(x, y, t) = g_A\, \theta(x_b(t) - x) + g_B\, \theta(x - x_b(t)),
\end{equation}

where $x_b(t)$ is the time-dependent position of the boundary, and $\theta(x)$ is the Heaviside step function.

This observation—that the vertical interface remains stable—motivates the following ansatz for the velocity field:

\begin{equation}
    \vec{v} =
    \begin{pmatrix}
        v_x(x,t) \\
        v_y(y,t)
    \end{pmatrix},
\end{equation}

i.e., the cross-derivatives vanish: $\partial_x v_y = \partial_y v_x = 0$. While the growth rate is position-dependent (and one strain grows faster than the other), this velocity field remains consistent with the observation because the faster-growing strain compresses the slower-growing one, effectively squeezing it out.

Solving the continuity equation $\nabla \cdot \vec{v} = g(x,y,t)$ with the above ansatz yields:

\begin{align}
    v_x &= \frac{g_B - g_A}{L} \left[ (x_b - L) x \theta(x_b - x)
    + x_b (x - L) \theta(x - x_b) \right], \\
    v_y &= g_{\rm average}  y,
\end{align}

where the average growth rate across the entire system is defined as:

\begin{equation}
    g_{\rm average}(t) = \frac{g_A\, x_b(t) + g_B\, (L - x_b(t))}{L}.
\end{equation}

The motion of the boundary is determined by the horizontal velocity at the boundary itself, i.e.,

\begin{equation}
    \dot{x}_b = v_x(x_b).
\end{equation}

Since the fraction of strain 1 is given by $c_A = \frac{x_b(t)}{L}$, we can rewrite the boundary motion equation as:

\begin{equation}
    \dot{c}_A = (g_A - g_B)\, c_A (1 - c_A),
\end{equation}

which is the classical logistic equation, describing the selective advantage of the faster-growing strain.

\section{Boundary shape} 
\label{note:Note5}

In the case where a small geometric defect is present at the floor of the cavity—but does not significantly perturb the overall flow field—we can assume that, far from the defect, the flow remains approximately the same as in the case of a cavity with a smooth floor. Specifically, the velocity field can be written as:

\begin{align}
    v_x(x, y) &= \frac{g_B - g_A}{L} \left[ (x_b - L) x \theta(x_b - x) + x_b (x - L) \theta(x - x_b) \right], \\
    v_y(y) &= g_{\rm average}\, y,
\end{align}

where $x_b(y)$ is the horizontal position of the strain boundary at height $y$, and $g_{\rm average}$ is the average growth rate across the cross-section at that height.

The velocity components evaluated at the strain boundary are:

\begin{align}
    v_x^b &= \frac{g_B - g_A}{L} (x - L)\, x, \\
    v_y^b &= g_{\rm average}\, y.
\end{align}

If the boundary $y_b(x)$ is stable, then the local velocity must be directed tangentially along the boundary. This implies:

\begin{equation}
    \frac{dy_b}{dx} = \frac{v_y^b}{v_x^b}.
\end{equation}

Substituting the expressions for $v_y^b$ and $v_x^b$ into this relation yields:

\begin{equation}
    \frac{dy_b}{dx} = \frac{g_{\rm average}\, y}{\frac{g_B - g_A}{L} (x - L)\, x}.
\end{equation}

Solving this differential equation gives the shape of the boundary:

\begin{equation}
    y_b(x) = C \left( \frac{x}{L - x} \right)^{\frac{g_{\rm average}}{g_A - g_B}},
    \label{eq:boundary-gen}
\end{equation}

where $C$ is an integration constant set by the boundary condition near the defect.

One can show that if the interstrain boundary is stable near the cavity floor, it must be perpendicular to the floor. We can prove this as follows.

For the purposes of this proof (but not for the rest of the manuscript), we redefine the coordinate system such that the origin is placed at the intersection point between the interstrain boundary and the cavity floor. Let the $X$-axis run along the floor, and the $Y$-axis be perpendicular to it, pointing into the cavity.

Since  the flux through the floor is zero we have  $v_y(x, 0) = 0$ in the vicinity of the intersection. At intersection itself we also have $v_x(0, 0) = 0$ since the boundary is stable.

Recalling that the velocity field is the gradient of the pressure, $\vec{v} = \nabla p$, we expand the pressure field $p(x, y)$ near the origin. To leading order, we obtain:

\begin{align}
    v_x &= \frac{\partial p}{\partial x} = p''_{xx}\, x, \\
    v_y &= \frac{\partial p}{\partial y} = p''_{yy}\, y,
\end{align}
where $p''_{xx}$ and $p''_{yy}$ denote second derivatives of $p$ evaluated at the origin.

This implies that $v_x(x = 0, y) = 0$ near the floor. Hence, there is no horizontal flow at the intersection, and the strain boundary near the origin will locally align with the $Y$-axis. In other words, the boundary must be perpendicular to the cavity floor.

Returning to the original coordinate system, we can use the perpendicularity condition to determine the integration constant in Eq.~\ref{eq:boundary-gen}. Suppose the floor surface is described by a function $y = f(x)$. Then, the point $(x^*, y^*)$ at which the interstrain boundary intersects the floor must satisfy two conditions:

\begin{align}
    y_b(x^*) &= f(x^*), \\
    y_b'(x^*)\, f'(x^*) &= -1.
\end{align}

The first condition ensures that the boundary touches the floor, while the second enforces that the boundary is perpendicular to the floor at the intersection point.

Differentiating the boundary shape from Eq.~\ref{eq:boundary-gen},
\[
y_b(x) = C \left( \frac{x}{L - x} \right)^{1 / \epsilon}, \quad \text{with } \epsilon = \frac{g_A - g_B}{g_{\rm average}},
\]
we obtain:
\[
\frac{d}{dx} \log y_b(x) = \frac{1}{\epsilon} \left( \frac{1}{x} + \frac{1}{L - x} \right).
\]

Substituting into the perpendicularity condition and expressing it in logarithmic derivative form, we get:

\begin{equation}
    \left( \log y_b \right)' = -\frac{2}{\left( f^2 \right)'},
\end{equation}

Consider the case where the cavity floor contains a smooth elliptic defect, centered at position $x_0$. The profile of the floor is then given by:

\begin{equation}
    f^2(x) = b^2 - \frac{b^2}{a^2}(x - x_0)^2,
\end{equation}

where $a$ and $b$ are the semi-major and semi-minor axes of the ellipse, respectively. Using the condition for a perpendicular intersection of the strain boundary and the floor, we substitute this into the earlier result:

\begin{equation}
    \frac{g_{\rm average}}{g_A - g_B} \left( \frac{L}{x} + \frac{L}{L - x} \right) = \frac{a^2}{b^2} \cdot \frac{1}{x - x_0}.
\end{equation}

To simplify the notation, define:

\begin{equation}
    z = \frac{b^2}{a^2} \cdot \frac{g_{\rm average}}{g_A - g_B}.
\end{equation}

Solving the resulting quadratic equation yields the position of the strain boundary:

\begin{equation}
    x_b = L \cdot \frac{\sqrt{(z - 1)^2 + 4z\, \frac{x_0}{L}} - (z - 1)}{2}.
\end{equation}

For the solution to be valid, the boundary must intersect the defect within its horizontal extent:

\begin{equation}
    x_0 - a < x_b < x_0 + a.
\end{equation}

This leads to the condition:

\begin{equation}
    z > \frac{L - x_0 - a}{L} \cdot \frac{x_0 + a}{a}.
\end{equation}

For small defects (i.e., $a \ll x_0, L - x_0$), this simplifies to:

\begin{equation}
    z > \frac{L - x_0}{L} \cdot \frac{x_0}{a}.
\end{equation}

Substituting back for $z$, we get:

\begin{equation}
    \frac{b^2}{a^2} \cdot \frac{g_{\rm average}}{g_A - g_B} > \frac{L - x_0}{L} \cdot \frac{x_0}{a},
\end{equation}

or, after rearranging:

\begin{equation}
    \frac{b}{a} \cdot b \cdot g_{\rm average} > \frac{g_A - g_B}{L} (L - x_0)\, x_0.
\end{equation}

Using the definitions of local velocity components, this can be interpreted as:

\begin{equation}
    \frac{b}{a} v_y > v_x,
    \label{eq:phys}
\end{equation}

which has a clear physical meaning: the vertical velocity generated by average growth (modulated by defect slope) must exceed the horizontal velocity driven by strain competition. In other words, the "upward" flow due to proliferation must outcompete the "sideways" push caused by the growth rate difference between strains in order for the defect to support a vertical boundary.

For the special case of a circular defect, $a = b = r$, and the condition simplifies to:

\begin{equation}
    r\, g_{\rm average} > \frac{g_A - g_B}{L} (L - x_0)\, x_0,
\end{equation}

which leads directly to Equation~\ref{eq:circular} in the main text.

\section{Jammed Population in Flowing Water} 
\label{note:Note6}

Even in the jammed state, where no additional space is available for more cells, nutrients must still be delivered to the bottom of the cavity. This implies the presence of a nutrient-carrying water flow, characterized by a water current $\vec{j}_w$, alongside the cellular current $\vec{j}_c$. The continuity equations for the cells and the water are then:

\begin{align}
    \frac{\partial \rho_c}{\partial t} + \nabla \cdot \vec{j}_c &= g\, \rho_c, \\
    \frac{\partial \rho_w}{\partial t} + \nabla \cdot \vec{j}_w &= -g\, \rho_c.
\end{align}

In a stationary state (i.e., when the densities are time-independent), these reduce to:

\begin{align}
    \nabla \cdot \vec{j}_c &= g\, \rho_c, \\
    \nabla \cdot \vec{j}_w &= -g\, \rho_c.
\end{align}

Thus, water is effectively consumed at a rate equal to the cell growth rate, to supply nutrients to proliferating cells.

In this regime, the densely packed microbial colony behaves as a porous medium through which water flows. According to Darcy's law~\cite{Darcy1856, Whitaker1986}, the relative velocity of the water with respect to the cells satisfies:

\begin{equation}
    \vec{v}_w - \vec{v}_c = -k\, \nabla p_w,
\end{equation}

where $k$ is the permeability (porosity-related coefficient) of the medium, and $p_w$ is the water pressure.

Cells experience two primary forces: (i) viscous friction due to the relative motion with respect to the water, proportional to $\vec{v}_c - \vec{v}_w$, and (ii) mechanical stress from neighboring cells. The force balance (modifying the Stokes equation~\ref{eq:Stokes}) is then:

\begin{equation}
    \gamma (\vec{v}_c - \vec{v}_w) = \nabla \cdot \tilde{\sigma} - \nabla p_c,
\end{equation}

where $\gamma$ is the friction coefficient between the cells and water, $\tilde{\sigma}$ is the deviatoric part of the cellular stress tensor (excluding isotropic pressure), and $p_c$ is the cellular pressure.

Substituting Darcy's law into this force balance yields:

\begin{equation}
    \nabla \cdot \tilde{\sigma} - \nabla \left( p_c + \gamma k\, p_w \right) = 0.
\end{equation}

This shows that the interaction with water effectively shifts the cellular pressure. Defining a modified total pressure $p = p_c + \gamma k\, p_w$, we recover the original form of the mechanical equilibrium equation (cf. Equations~\ref{eq:stress} and~\ref{eq:Stokes}).

The primary effect of water-cell interactions in the jammed state is a shift in the pressure field within the colony. While the absolute pressure measured inside the cavity will differ from the no-water case due to frictional coupling, the velocity field—and thus the structure of the flow and cell displacement—remains unchanged.

\section{Solution of the Stokes Equations for Growing Viscous Fluid} 
\label{note:Note7}

To solve the Stokes equations for the velocity field inside the cavity, we begin by writing them explicitly in components:

\begin{align}
    \eta (\partial_x^2 + \partial_y^2)\, v_x &= \partial_x p, \\
    \eta (\partial_x^2 + \partial_y^2)\, v_y &= \partial_y p,
\end{align}

where $\eta$ is the viscosity and $p$ is the pressure. The incompressibility condition, modified by the growth rate, is given by:

\begin{equation}
    \partial_x v_x + \partial_y v_y = g(x),
\end{equation}

In this section we denote $L$ the half width of the cavity and place the origin of the coordinate to the middle of the floor, what simplifies boundary conditions:

\begin{equation}
    \eta \partial_x v_y(\pm L) \pm \xi\, v_y(\pm L) = 0,
\end{equation}

where $\xi$ is the effective wall friction coefficient.

For the two-strain configuration forming vertical columns (as observed in simulations), the growth rate is given by:

\begin{equation}
    g(x) = g_A\, \theta(x_b - x) + g_B\, \theta(x - x_b),
\end{equation}

where $x_b$ denotes the boundary between strains.

Motivated by numerical observations, we seek an analytical solution of the form:

\begin{equation}
    v_x = f(x), \quad v_y = y\, h(x),
\end{equation}

which satisfies the incompressibility condition:

\begin{equation}
    h(x) + f'(x) = g(x).
\end{equation}

Substituting into the Stokes equations gives:

\begin{align}
    \eta\, y\, h''(x) &= \partial_y p, \\
    \eta\, f''(x) &= \partial_x p.
\end{align}

The boundary conditions at $x = \pm L$ become:

\begin{equation}
    \eta\, h'(\pm L) \pm \xi\, h(\pm L) = 0.
\end{equation}

Solving the equations above yields:

\begin{align}
    h(x) &= \frac{H_2}{2} \left(x^2 - L^2 - 2 \frac{\eta}{\xi} L\right), \\
    f(x) &= g(x)\, (x - x_b) + F_0 - \frac{H_2}{2} \left( \frac{x^3}{3} - L^2 x - 2 \frac{\eta}{\xi} L x \right),
\end{align}

where $F_0$ and $H_2$ are constants determined by matching conditions. Specifically:

\begin{align}
    F_0 &= \frac{1}{2} \left[ (g_A + g_B) x_b - (g_A - g_B) L \right], \\
    H_2 \left( \frac{L^3}{3} + \frac{\eta}{\xi} L^2 \right) &= \frac{1}{2} \left[ (g_A - g_B) x_b - (g_A + g_B) L \right].
\end{align}

To ensure boundary stability, the horizontal velocity must vanish at the interstrain boundary: $v_x(x_b) = f(x_b) = 0$. Substituting into the expression for $f(x)$ gives:

\begin{equation}
    F_0 - \frac{H_2}{2} \left( \frac{x_b^3}{3} - L^2 x_b - 2 \frac{\eta}{\xi} L x_b \right) = 0.
\end{equation}

Solving this equation for $x_b$ and returning back to original coordinates we obtain Eq. \ref{eq:friction-x} in the main text. The other possible solutions are $x_b = 0,L$ what corresponds to only one strain existing. In this case, the velocity profile takes a simple parabolic shape that can be compared with the experiment (see Fig.~\ref{fig:friction}B in the main text)

\begin{eqnarray}
    v_y = v_0 \frac{x (L - x) + \frac{\eta}{\xi} L}{L \left(\frac{L}{6} + \frac{\eta}{\xi}\right) }.
    \label{eq:parabolic}
\end{eqnarray}

\end{document}